\documentclass[aps,prl,twocolumn,showpacs,subfigure,superscriptaddress,nobibnotes,nofootinbib]{revtex4-1}
\usepackage[dvips]{graphicx}
\usepackage{epsfig}
\usepackage{bm}   
\usepackage{dcolumn}
\usepackage{graphicx}
\usepackage{rotate}
\usepackage{amsfonts}
\usepackage[table]{xcolor}
\usepackage{tabularx}
\usepackage{footmisc,booktabs,amssymb}

\begin{document}

\title{New formulae for the $(-2)$ moment of the photo-absorption cross section, $\sigma_{_{-2}}$}

\author{J. N.~Orce}
\email{coulex@gmail.com} \homepage{http://www.pa.uky.edu/~jnorce}
\affiliation{Department of Physics, University of the Western Cape, P/B X17, Bellville, ZA-7535 South Africa}

\date{\today}

\begin{abstract}

Two new  formulae for the $(-2)$ moment of the photo-absorption cross section, $\sigma_{_{-2}}$, 
have  been determined, respectively, from  the 1988  compilation of Dietrich and Berman and a mass-dependent 
symmetry energy coefficient, $a_{sym}(A)$. 
The data for $A\gtrsim50$ follow, with a {\small $RMS$} deviation of 6\%, the power-law $\sigma_{_{-2}}=2.4 A^{5/3}$$\mu$b/MeV,  
which  is in agreement with Migdal's calculation of 
$\sigma_{_{-2}}=2.25A^{5/3}$$\mu$b/MeV based on the hydrodynamic model and the $\sigma_{_{-2}}$ sum rule.
The additional inclusion of $a_{sym}(A)$ provides a deeper insight to the nuclear polarization of $A\geqq10$ nuclei.  
\end{abstract}

\pacs{21.10.Ky,  25.70.De,  25.20.-x, 25.20.Dc, 24.30.Cz}

\keywords{photo-absorption cross section, nuclear polarizability, spectroscopic quadrupole moment}

\maketitle


The ratio of the induced dipole moment to an applied constant electric field 
yields the static nuclear polarizability, $\alpha$. 
On using the hydrodynamic model and assuming  inter-penetrating 
proton and neutron fluids with a well-defined nuclear surface of radius $R=r_{_0}A^{1/3}$ fm, 
Migdal~\cite{migdal,levinger} obtains, 
\begin{equation}
\alpha=\frac{e^2R^2A}{40~a_{sym}}=2.25\times 10^{-3} A^{5/3} \mbox{fm}^3, 
\label{eq:sigma-2}
\end{equation}
where $a_{sym}=23$ MeV is the symmetry energy coefficient in the Bethe-Weizs\"acker semi-empirical  mass formula~\cite{weiz,bethe} 
and $r_{_0}=1.2$ fm.
This semiclassical treatment  
considers  the nuclear symmetry energy,  $a_{sym}(N-Z)^2/A$, 
to be spread  uniformly throughout the nucleus as a symmetry energy density $a_{sym}(\rho_n-\rho_p)^2/\rho$.

Alternatively,  $\alpha$  can be calculated from the $(-2)$ moment of the total electric-dipole 
photo-absorption cross section, $\sigma_{_{-2}}$, 
\begin{eqnarray}
\sigma_{_{-2}}&=&\int_0^\infty\frac{\sigma_{_{total}}(E_{_{\gamma}})}{E_{_{\gamma}}^2} ~dE_{_{\gamma}},
\label{eq:sigma-222}
\end{eqnarray}
using second-order perturbation theory~\cite{levinger2,migdal3}. It follows from the sum rule\footnote{The dimensionless 
oscillator strength $f_{in}$ for $E1$ transitions,
\begin{equation}
 f_{in}=\frac{2M}{\hbar^2}E_{_{\gamma}}\langle i\parallel\hat{E1}\parallel n\rangle \langle n\parallel\hat{E1}\parallel i\rangle, 
\end{equation}
and its relation with the total photoabsorption cross section, 
\begin{equation}
 \int_0^\infty \sigma_{_{total}}(E_{_{\gamma}})dE_{_{\gamma}}=\frac{2\pi^2 e^2 \hbar}{Mc}\sum_n f_{in},
\end{equation}
are introduced in Eqs.~\ref{eq:polar} and \ref{eq:fsigma}, respectively~\cite{mott,merzbacher,levinger2}.}, 
\begin{eqnarray}
\alpha&=&2e^2\sum_n \frac{\langle i\parallel\hat{E1}\parallel n\rangle \langle n\parallel\hat{E1}\parallel i\rangle}{E_{_{\gamma}}} \label{eq:polar} 
\\ 
&=&\frac{e^2\hbar^2}{M}\sum_n \frac{f_{in}}{E_{_{\gamma}}^2} = \frac{\hbar c}{2\pi^2}\int_0^\infty 
\frac{\sigma_{_{total}}(E_{_{\gamma}})}{E_{_{\gamma}}^2} ~dE_{_{\gamma}} \label{eq:fsigma} \\ 
&=& \frac{\hbar c}{2\pi^2}\sigma_{_{-2}},
\label{eq:sigma-22}
\end{eqnarray}
where $E_{_{\gamma}}$ is the $\gamma$-ray energy corresponding to a transition connecting the ground state $|i\rangle$ 
and an excited state $|n\rangle$, $M$  the nucleon mass, 
and $\sigma_{_{total}}(E_{_\gamma})$ the total photo-absorption cross section. The $\sigma_{_{total}}(E_{_\gamma})$  
cross section generally includes the $(\gamma,n)  +  (\gamma,p) + (\gamma,np) + (\gamma,2n) + (\gamma,3n) + (\gamma,F)$ channels, 
which are in competition in the giant dipole resonance ({\small GDR}) region~\cite{GDRreview,lectures}. 

\begin{figure}[!h]
\begin{center}
\includegraphics[width=7.5cm,height=7cm,angle=-0]{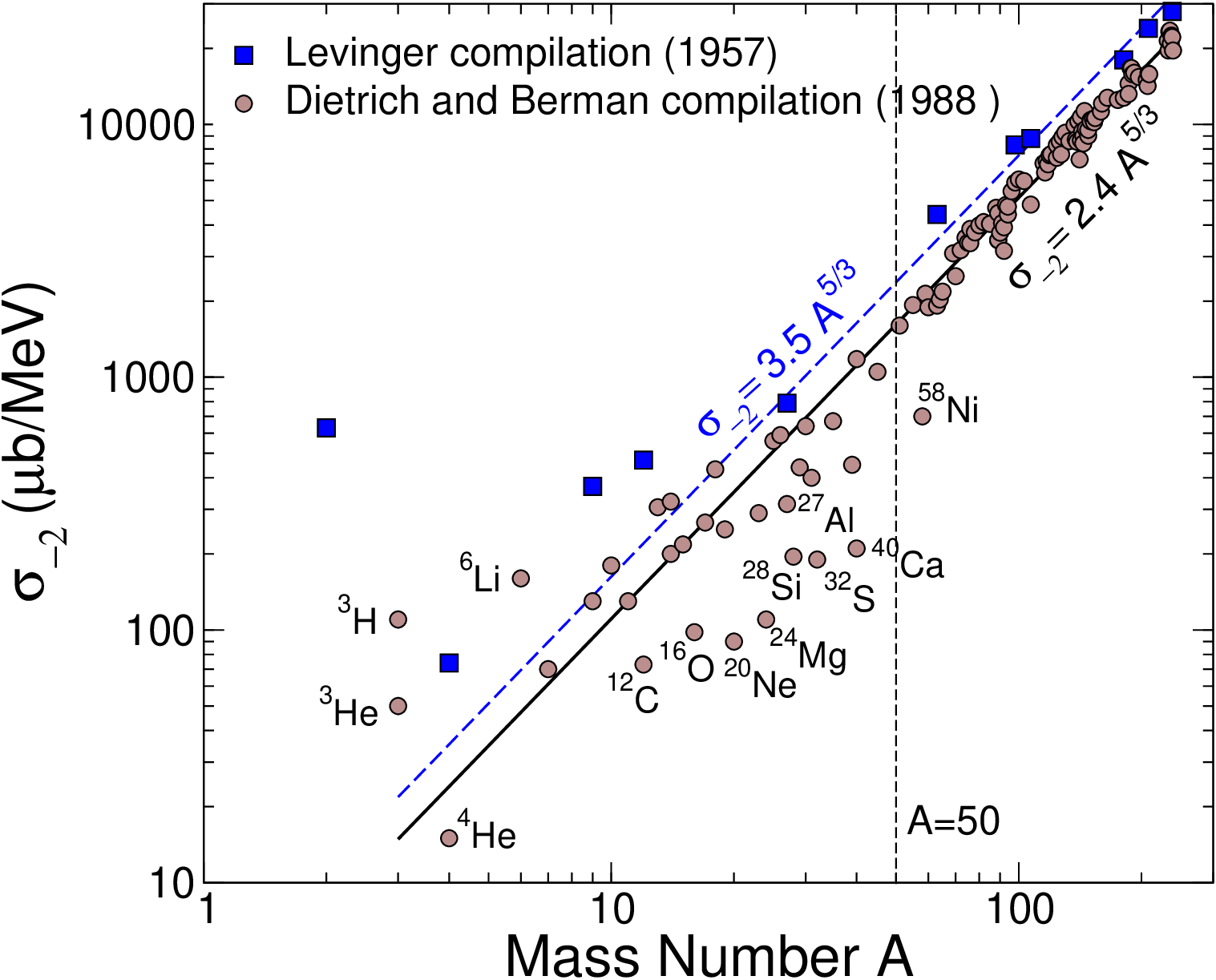} 
\caption{(Color online) The $(-2)$ moment of the total photo-absorption cross section $\sigma_{_{-2}}$ $vs$ A on a log-log scale. 
The experimental values from the 1988 
compilation~\cite{atlas} are given by circles. These data follow a power-law relationship  $\sigma_{_{-2}}=2.4$A$^{5/3}$ $\mu$b/MeV. 
The dashed line represents Levinger's fit to the available data (squares) in 1957, $\sigma_{_{-2}}=3.5$A$^{5/3}$ $\mu$b/MeV~\cite{levinger}. 
In both cases, $\kappa=1$ is assumed.} 
\label{fig:sigma}
\end{center}
\end{figure}

On comparing Eqs.~\ref{eq:sigma-2} and \ref{eq:sigma-22}, Migdal extracted $\sigma_{_{-2}}$ as~\cite{migdal} 
\begin{equation}
\sigma_{_{-2}}=2.25 A^{5/3} \mbox{$\mu$b/MeV}.\label{eq:migdal}
\end{equation}
This power-law relationship was empirically confirmed by Levinger in 1957 from a fit to the  
available $\sigma_{_{-2}}$ data~\cite{levinger},  
\begin{equation}
\sigma_{_{-2}}=3.5 \kappa A^{5/3} \mbox{$\mu$b/MeV}. 
\label{eq:3p5}
\end{equation}
Levinger's fit is shown in Fig.~\ref{fig:sigma} and included eleven $\sigma_{_{-2}}$ data points (squares) 
with approximate estimations for the high-energy, 
neutron multiplicity and $\sigma(\gamma,p)$ contributions. 
The polarizability parameter $\kappa$ is the ratio of the observed {\small GDR}  effect to that predicted by the 
hydrodynamic model~\cite{levinger}, as determined by comparing the measured $\sigma_{_{-2}}$ values and Eq.~\ref{eq:3p5}.
This comparison yields $\kappa=1$ for the ground state of nuclei with  $A\gtrsim20$~\cite{levinger}.  
Lighter nuclei require larger values of $\kappa$ to reproduce the data.
Using Eqs.~\ref{eq:sigma-22} and \ref{eq:3p5}, 
the nuclear polarizability is given by
\begin{eqnarray}
   \alpha=3.5k\times 10^{-3} \mbox{A}^{5/3} ~\mbox{fm}^3, \label{eq:alpha}
\end{eqnarray}
which depends on the nuclear size and $\kappa$.


%

In 1988, Dietrich and Berman re-compiled the  photoneutron  cross-section data~\cite{atlas}. 
This compilation included  $(\gamma,n) + (\gamma,pn) + (\gamma,2n) + (\gamma,3n) + (\gamma,F)$ 
data from studies at Livermore, Giessen, Saclay and other laboratories 
which used monochromatic photon beams generated by in-flight annihilation of positrons\footnote{Most of the photonuclear 
data produced during 1960-1988 was taken with monochromatic photon beams~\cite{lectures,atlas}. One main advantage of 
this technique over bremsstrahlung photon beams, broadly used prior to 1960, is the 
direct and simultaneous measurements of the partial photoneutron 
cross sections which are in competition in the {\small GDR} region. 
These simultaneous measurements are essential to obtain a reliable $\sigma_{_{total}}(E_{_\gamma})$~\cite{lectures}.}.

Figure~\ref{fig:sigma} shows the $\sigma_{_{-2}}$ data (in $\mu$b/MeV) from the Dietrich and Berman compilation (circles) ~\cite{atlas}, 
by integrating Eq.~\ref{eq:sigma-222} 
between the $(\gamma,n)$ threshold and an upper limit of  ${E_{\gamma_{max}}}\approx20-50$ MeV. These integration limits
include the {\small GDR} but do not take into consideration $\sigma(\gamma,p)$  contributions and the 
rise of $\sigma(E_{_{\gamma}})$ at around 140 MeV\linebreak due to pion exchange currents~\cite{320MeV}. 
Because of the $1/E_{_{\gamma}}^2$ factor,  $\sigma_{_{-2}}$ is less sensitive to these 
high-energy contributions, which account for less than 10\% of the total $\sigma_{_{-2}}$ value~\cite{levinger,320MeV,kerst_price,ahrens}. 
This plot uses the mean value  when several measurements were available for the same isotope and 
excluded data from natural samples unless one single isotope dominated the isotopic abundance. 

\begin{figure}[!ht]
\begin{center}
\includegraphics[width=7.cm,height=5.5cm,angle=-0]{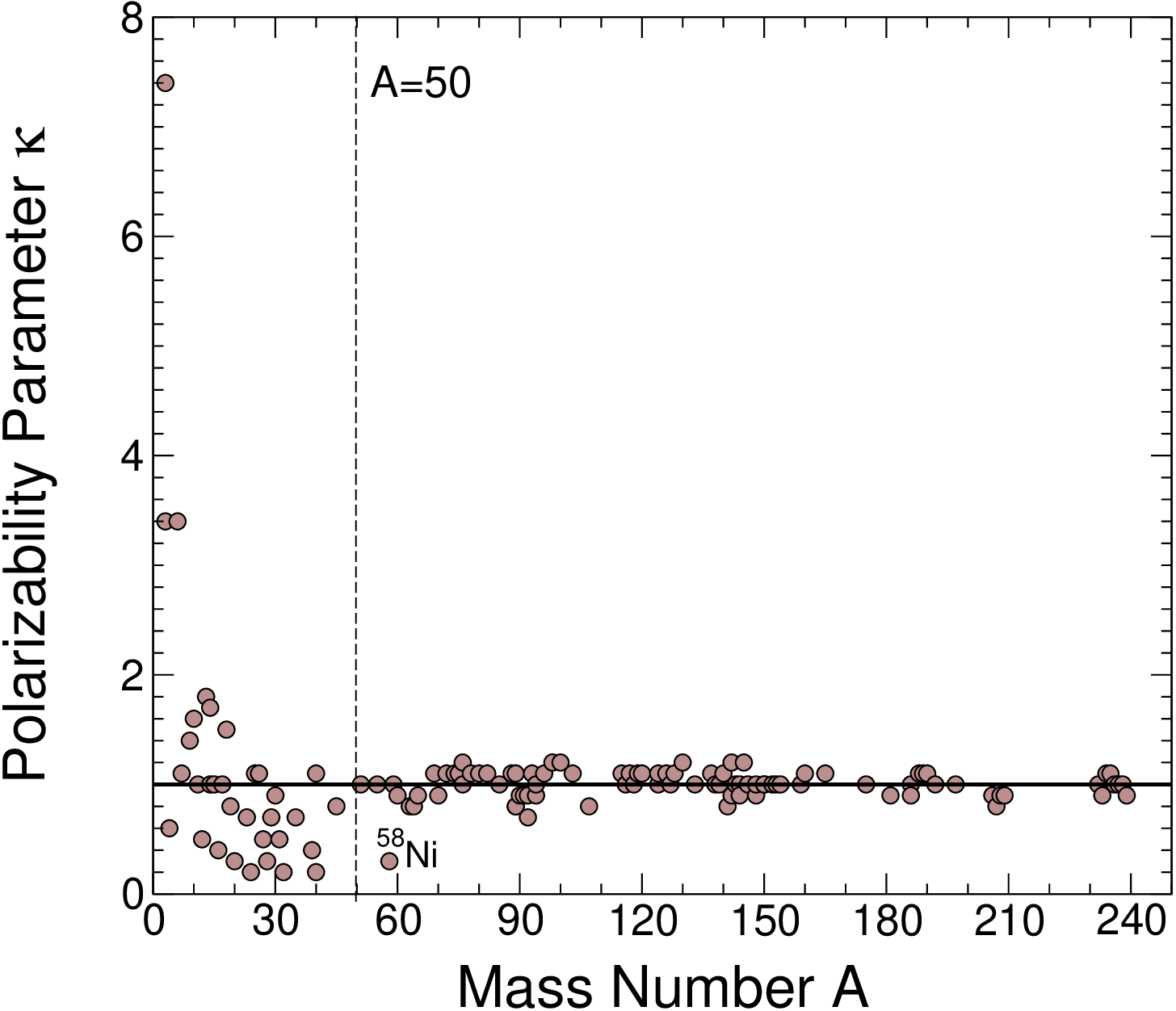} 
\caption{(Color online) The polarizability parameter $\kappa$ given by $\frac{\sigma_{_{-2}}}{2.4A^{5/3}}$ using the  
$\sigma_{_{-2}}$ data in the Dietrich and Berman compilation (circles)~\cite{atlas}. The horizontal solid line corresponds to $\kappa=1$. 
Large deviations from the hydrodynamic model prediction ($\kappa=1$) are observed for $A\lesssim50$.} 
\label{fig:polarizability}
\end{center}
\end{figure}

These data follow a power-law,
\begin{equation}
\sigma_{_{-2}}=2.4\kappa A^{5/3}~{\rm \mu b/MeV},
\label{eq:mine}
\end{equation}
with a {\small $RMS$} deviation of 30\% for $\kappa=1$. 
For $A\geqq50$, on excluding the $^{58}$Ni data point which has a large $\sigma(\gamma,p)$ contribution~\cite{berman1975,lectures2}, 
the agreement is even better, as shown in Fig.~\ref{fig:sigma}, 
with a {\small $RMS$} deviation of 6\%. 
This formula agrees with the one published by Berman 
and Fultz in their 1975 review paper for $A\gtrsim60$: $\sigma_{_{-2}}=2.39(20)A^{5/3}$$\mu$b/MeV~\cite{berman1975}.
For $A<50$, Fig.~\ref{fig:sigma} presents large deviations from $\kappa=1$ for $A=4n$, $T_{_Z}=0$ nuclei ($\kappa<1$) and 
loosely-bound light nuclei with $A<20$ ($\kappa>1$). To emphasize this point, Fig.~\ref{fig:polarizability} shows a similar 
plot of the polarizability parameter $\kappa~vs.~A$ by comparing Eq.~\ref{eq:mine} and the empirical $\sigma_{_{-2}}$ values~\cite{atlas}.


The missing $\sigma(\gamma,p)$ contribution in the Dietrich and Berman compilation is 
the reason for the $\kappa<1$ values observed\footnote{The total photoneutron 
cross section, $\sigma(\gamma,n)$, for $^{58}$Ni is relatively small 
because of the $\frac{\sigma(\gamma,p)}{\sigma(\gamma,n)}$ ratio is also 
controlled by the relative level densities in the residual nuclei, i.e., 
the ratio of the number of open channels, $\frac{N_p}{N_n}$. 
For $^{58}$Ni, $\frac{N_p}{N_n}\approx\frac{\sigma(\gamma,p)}{\sigma(\gamma,n)}=2$~\cite{lectures2}.}
for many $A<50$ nuclei and $^{58}$Ni. 
For heavier nuclei, neutron emission is the favorable decay mode due to the strong suppression 
of proton emission by the Coulomb barrier. 
Proton emission is, however, the predominant decay mode for $A=4n$ self-conjugate nuclei with $A\lesssim50$~\cite{morinaga2,sven}. 
For example,  $\sigma(\gamma,p)\approx7\times\sigma(\gamma,n)$ in $^{40}$Ca~\cite{balashov}. 
This is because of the  isospin selection rule $\Delta T=\pm 1$ for $E1$ excitations in a $T_{_Z}=0$ self-conjugate 
nucleus\footnote{$\Delta T=\pm1$ 
isovector transitions are isospin forbidden as the Wigner coefficient $\left( \begin{array}{ccc} T_f &  1 &  T_i \\
-T_{_Z} &  0 & T_{_Z} \end{array} \right)$ vanishes for  $T_{_Z}=0$~\cite{isospin}.}.
For a nucleus with a ground state of isospin $T$, there is an isospin splitting of the {\small GDR}~\cite{shoda} which corresponds to  
excited proton ($T+1$) and neutron ($T$) resonances, with the $T+1$ resonance generally lying at a higher excitation energy. 
The isospin selection rule\footnote{With only small admixtures~\cite{barker2,kuhlmann}, isospin is 
a good approximation in photonuclear reactions for light nuclei involving photons in the range of the electric dipole 
absorption~\cite{gellmann}.} favors the excitation of $T+1$ states~\cite{isospin}, 
which predominantly decay by proton emission\footnote{Most neutron emission from excited states with isospin $T+1$ is forbidden, 
whereas neutron emission from excited states with isospin $T$ is allowed~\cite{morinaga3}. These selection rules 
follow from the respective Clebsch-Gordan coefficients in the transition probabilities.}~\cite{morinaga3}. 
Although the $\sigma(\gamma,p)$ data are scarce, the $\sigma_{_{-2}}$ sum rule~\cite{levingerbethe} seems to be exhausted 
once  the $\sigma(\gamma,p)$ contributions are included~\cite{sven,morinaga2,halpern}.\\



The larger {\small GDR} effect ($\kappa>1$) observed in Fig.~\ref{fig:polarizability} for light nuclei with $A\lesssim20$ may be explained from the 
mass dependence of the symmetry energy coefficient,  a$_{sym}(A)$, of relevance to test $3N$ forces~\cite{hebeler} 
and describe neutron stars and supernova cores~\cite{latimer,pearson}. 
As mentioned above, Migdal utilized a constant value of $a_{sym}=23$ MeV to determine $\sigma_{_{-2}}$ in Eq.~\ref{eq:migdal}. 
Nevertheless, the mass dependence of $a_{sym}(A)$ has long been established in the liquid droplet model~\cite{droplet} 
and recognized as the fundamental parameter describing the {\small GDR}~\cite{berman1975}. 
Its form has since been refined, despite its current model dependency~\cite{tian}, with the advent of high-precision  
mass measurements.

\begin{figure}[!h]
\begin{center}
\includegraphics[width=6.5cm,height=6.cm,angle=-0]{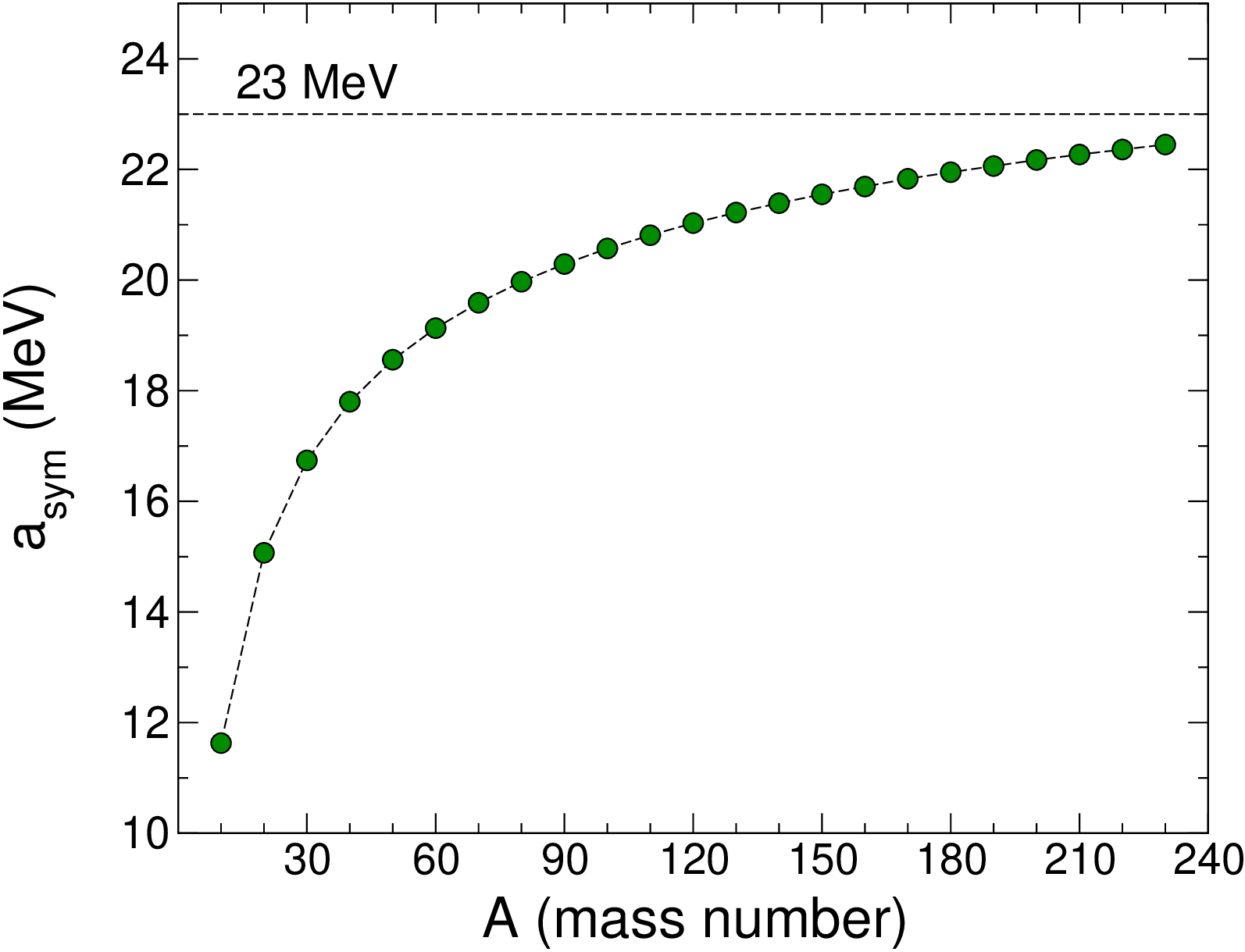} \\
\caption{(Color online) Symmetry energy coefficient, $a_{sym}(A)$, of finite nuclei  as a function of mass number $A$ using  
Eq.~\ref{eq:asym}~\cite{tian}.}
\label{fig:asym}
\end{center}
\end{figure}

From a global fit to the binding energies of isobaric nuclei with $A\geq10$~\cite{tian}, 
extracted from the 2012 atomic mass evaluation~\cite{audi}, Tian and co-workers determined 
a$_{sym}(A)$ as, 
\begin{equation}
 a_{sym}(A)=S_v\left(1-\frac{S_s}{S_vA^{1/3}}\right),
 \label{eq:asym}
\end{equation}
with $S_v\approx 28.32$ MeV being the bulk symmetry energy coefficient 
and $\frac{S_s}{S_v}\approx 1.27$ the surface-to-volume ratio\footnote{Similar coefficients are determined 
in Ref.~\cite{dieperink}.}.  
Within this approach,  the extraction of $a_{sym}(A)$ only  depends on the Coulomb energy term in the 
Bethe-Weizs\"acker semi-empirical  mass formula and shell effects~\cite{massmodel}, which are both 
included in Eq.~\ref{eq:asym}~\cite{tian}. 
Figure~\ref{fig:asym} illustrates the mass dependency of 
$a_{sym}(A)$ and clearly prevents the use of a constant $a_{sym}$ value. 

\begin{figure}[!h]
\begin{center}
\includegraphics[width=7.5cm,height=7.cm,angle=-0]{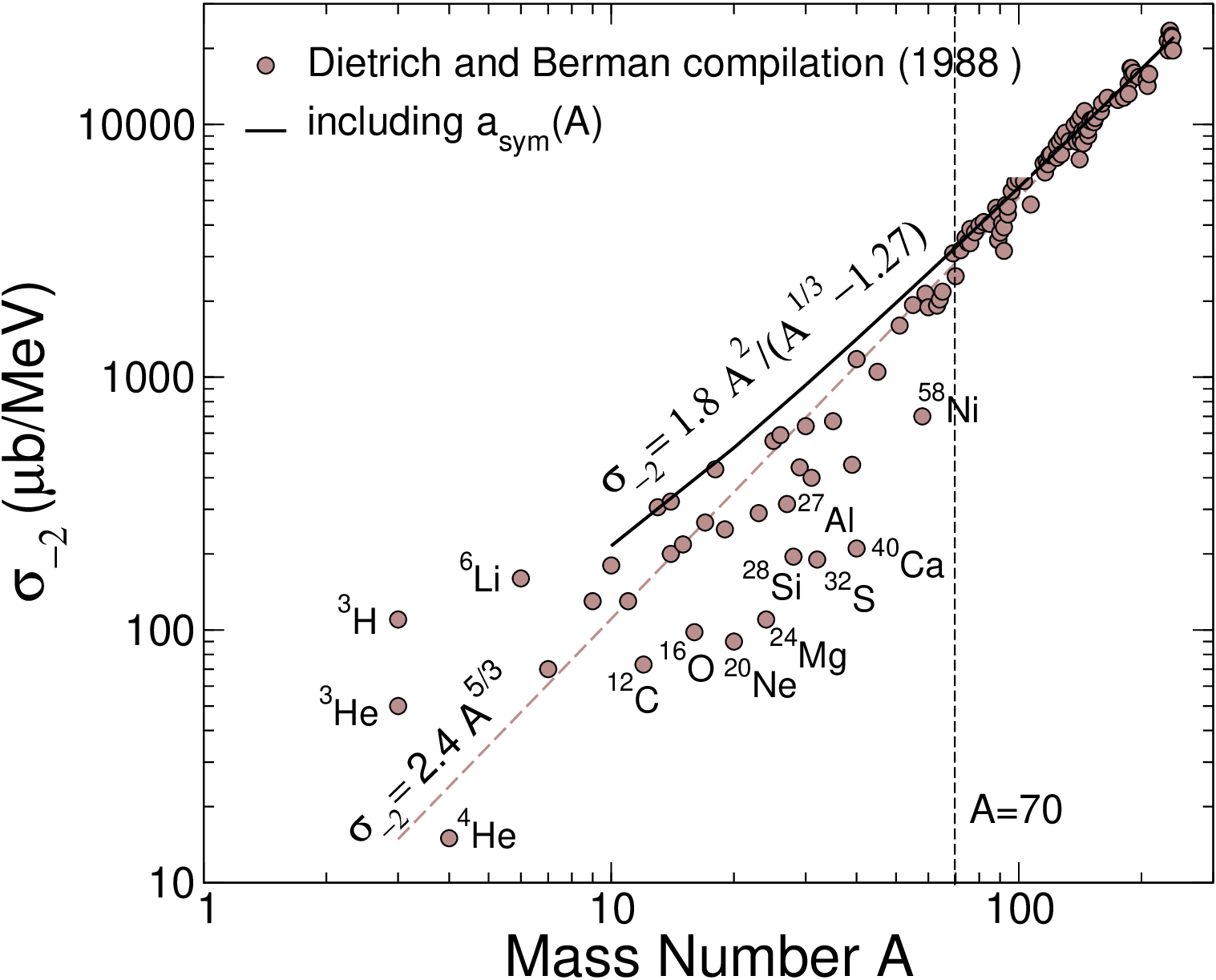} 
\caption{(Color online) The $(-2)$  moment of the total photo-absorption cross section $\sigma_{_{-2}}$ $vs$ A on a log-log scale 
using Eq.~\ref{eq:mine2} (solid line). For comparison purposes, Eq.~\ref{eq:mine} (dashed line) 
and the data from the 1988 compilation~\cite{atlas} are also plotted.}
\label{fig:sigmasym}
\end{center}
\end{figure}

After introducing this mass dependence in Eqs.~\ref{eq:sigma-2} and \ref{eq:sigma-22}, $\alpha$ and $\sigma_{_{-2}}$ are 
given by, 
\begin{equation}
 \alpha=\frac{1.8\times10^{-3} A^2}{A^{1/3}-1.27}~\mbox{fm}^3, 
 \label{eq:mine3}
\end{equation}
\begin{equation}
\sigma_{_{-2}}=\frac{1.8 A^{2}}{A^{1/3}-1.27} ~\mu\mbox{b/MeV}.
\label{eq:mine2}
\end{equation}
Equation \ref{eq:mine2} is plotted in Fig.~\ref{fig:sigmasym} for $A\geq10$ nuclides (solid line). 
Encouragingly, the increasing upbend observed as $A$ decreases  provides an explanation for the large  
{\small GDR} effects observed in light nuclei. However, the validity of the hydrodynamic model 
remains to be tested for the lightest $A<10$ nuclei. 

More generally, Eq.~\ref{eq:mine2} provides a means to evaluate  nuclear polarizability without invoking a polarizability parameter. 
As shown in Fig.~\ref{fig:sigmasym}, most of the data points either fall below the predicted curve  ($A<70$) 
or merge with it  where neutron emission is favorable ($A\geq70$). 
These facts indicate that  Eq.~\ref{eq:mine2} could exhaust the $\sigma_{_{-2}}$ sum rule for  
%
both photoneutron and photoproton cross sections and, hence,  incorporate 
the actual {\small GDR} effect to the nuclear polarizability. 
Consequently, the mass-dependent $\sigma_{_{-2}}$ curve may provide an estimate for the missing $\sigma(\gamma,p)$ contribution. 
For example,  the predicted value of $\sigma_{_{-2}}$ for $^{40}$Ca is in agreement with the experimentally determined 
$\sigma(\gamma,p)/\sigma(\gamma,n)$ ratio~\cite{balashov}. 
Additional experimental and theoretical work are needed
to test the generality of these findings and evaluate deviations from the hydrodynamic model. 


In conclusion, a new empirical formula (Eq.~\ref{eq:mine}) for the $(-2)$ moment of the photo-absorption cross section, $\sigma_{_{-2}}(A)$, 
has been determined from the latest photoneutron cross-section compilation with monoenergetic photons. 
The $\sigma_{_{-2}}$ data 
include most of the photoneutron channels but excludes relevant $\sigma(\gamma,p)$ contributions for $A\lesssim50$ nuclides. 
This new empirical formula presents a {\small $RMS$} deviation of 6\% for $A\gtrsim50$ and is 
in better agreement with Migdal's calculation of $\sigma_{_{-2}}$ (Eq.~\ref{eq:migdal}) 
on combining the hydrodynamic model and second-order perturbation theory.

Additionally, $\sigma_{_{-2}}$ has been inferred (Eq.~\ref{eq:mine2}) using a mass-dependent symmetry 
energy coefficient, $a_{sym}(A)$, determined by Tian and collaborators for $A\geq10$ nuclei, 
which includes Coulomb energy and shell corrections. 
The resulting curve seems to account for the actual {\small GDR} effects as it exhausts the $\sigma_{_{-2}}$ sum rule for 
most $A\geq10$ nuclei in the Dietrich and Berman compilation. Moreover, it provides an explanation for the larger 
polarization effects found in light nuclei with $10\leq A\lesssim20$. 
Additional work is needed to test this new equation and evaluate deviations from the hydrodynamic model.  
It is encouraging, though, that the curve nicely merges with the  $\sigma_{_{-2}}$ data for $A\gtrsim70$, in agreement with 
the dominant photoneutron cross sections for heavy nuclei. 

Data compilations of currently available photoproton and photoneutron cross sections remain to be done. 
The $\sigma(\gamma,p)$ data are scarce compared to the $\sigma(\gamma,n)$ data and extensive work is desirable  
throughout the nuclear chart. These new data are crucial to test the $\sigma_{_{-2}}$ sum rule and provide a means to 
remove the model dependency of $a_{sym}(A)$, which, in turn, may lead to a better understanding of $3N$ forces, 
neutron stars and supernova cores. 

Furthermore, this work has direct implications in: 1) broadly used  Coulomb-excitation codes such as {\small GOSIA}~\cite{gosia}, 
where the polarization potential has to be modified with either Eq.~\ref{eq:mine}, which requires a determination of $\kappa$ 
for $A\lesssim50$ nuclei, or Eq.~\ref{eq:mine2}, once its generality has been fully tested; and 2) 
shell model calculations of $\kappa$~\cite{barker,hausser2,orce}. To date, 
both approaches have broadly regarded Levinger's empirical formula (Eq.~\ref{eq:3p5}). 
For clarity purposes, these implications will be presented in a separate manuscript. 

The author acknowledges fruitful physics discussions with G. C. Ball, P. Navr\'atil, S. Triambak, D. H. Wilkinson and J. L. Wood. 
This work was supported by the South African National Research Foundation (NRF) under Grant 93500.

\end{document}